\documentclass[useAMS,usenatbib]{mn2e}
\usepackage{epsfig}
\usepackage{amssymb}
\usepackage{amsmath}
\usepackage{lscape}
\usepackage{longtable}
\usepackage[normal]{caption}
\usepackage{appendix}
\usepackage{bm}

\title[On the intrinsic AGN emission in far-infrared/sub-mm]{On the intrinsic AGN emission in the far-infrared/sub-mm}

\author[M.~Symeonidis et al.] 
{\parbox{\textwidth}{\raggedright
M.~Symeonidis,$^{1}$\thanks{E-mail: \texttt{m.symeonidis@ucl.ac.uk}}
}\vspace{0.4cm}\\
\parbox{\textwidth}{\raggedright $^{1}$ Mullard Space Science
  Laboratory, University College London, Holmbury St. Mary, Dorking,
  Surrey RH5 6NT, UK\\
}}

\begin{document}

\date{Accepted  Received; in original form}

\pagerange{\pageref{firstpage}--\pageref{lastpage}} \pubyear{2014}

\maketitle

\label{firstpage}

\begin{abstract}
Far-infrared (far-IR)/sub-mm emission linked to AGN-heated dust has been a topic of contention for many years. Results have been diverse and various views have been presented. The empirical AGN SED derived by Symeonidis et al. (2016, hereafter S16) has more far-IR/sub-mm emission than other SEDs in the literature, and thus it is contested by other works which argue that its luminosity in that part of the spectrum is overestimated. Here, I investigate this topic and the concerns raised over the S16 AGN SED. I also examine the differences between the S16 AGN SED and other commonly-used empirical AGN SEDs. My findings show that the reasons proposed by other works as to why the S16 AGN SED is not a reasonable representation of AGN emission in the far-IR/sub-mm, do not hold. 
\end{abstract}


\section{Introduction}
\label{sec:introduction}

\begin{figure*}
\epsfig{file=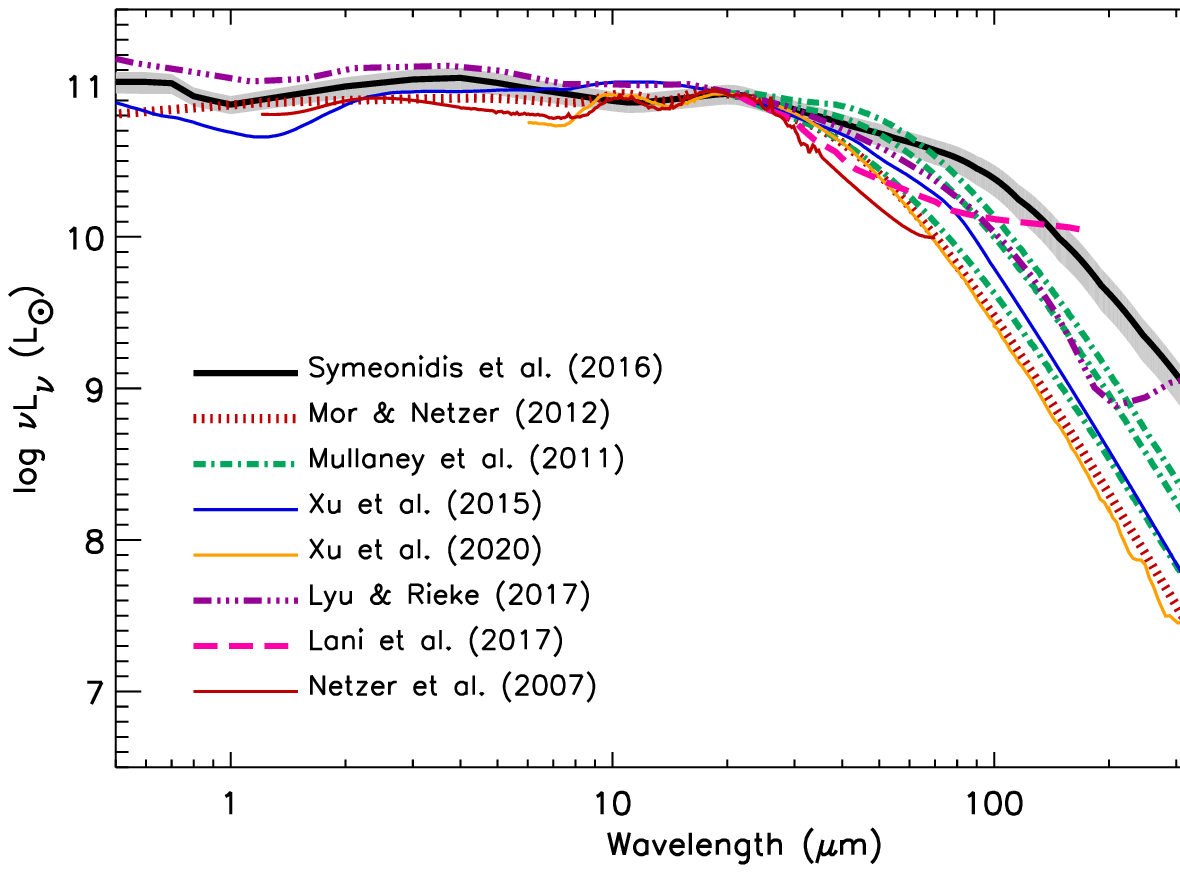,width=0.7\linewidth} 
\caption{The Symeonidis et al. (2016) intrinsic AGN SED (black solid curve) and 68 per cent confidence intervals (shaded region), compared with: (i) the AGN SEDs from Mullaney et al. (2011; dashed-dot green curves), (ii) the Xu et al. (2015) SED; solid blue curve, (iii) the extended Mor $\&$ Netzer (2012) SED; vertical-dash red curve, (iv) the Xu et al. (2020) SED; solid orange curve, (v) the Lyu $\&$ Rieke (2017) SED; dash-triple-dot purple curve, (vi) the Lani, Netzer $\&$ Lutz (2017) SED; dashed pink curve and (vii) the Netzer et al. (2007) SED; solid red curve. All SEDs are normalised to the S16 AGN SED at 20$\mu$m.  }
\label{fig:AGNseds}
\end{figure*}

The topic of far-IR emission linked to active galactic nucleus (AGN)-heated dust has been a topic of contention for many years. It is nevertheless particularly important to get right when aiming to disentangle the stellar and AGN components in galaxies' spectral energy distributions (SEDs), with far-reaching implications for computing galaxy star formation rates (SFRs) and understanding the relation between active black holes and their host galaxies. 

AGN have a large amount of dust in their vicinity (dust torus; e.g. Jaffe et al. 2004\nocite{Jaffe04}; Osterbrock $\&$ Ferland 2006\nocite{OF06}; Rodriguez-Ardila $\&$ Mazzalay 2006\nocite{RAM06}) which intercepts the UV and optical emission from the accretion disc re-radiating it in the near/mid-IR. The temperature of the dust is high ($\lesssim 2000$\,K; e.g. Barvainis et al. 1987\nocite{Barvainis87}; Schartmann et al. 2005\nocite{Schartmann05}) so it peaks in the 1-10$\mu$m region and declines rapidly longwards of $\sim$40$\mu$m. The AGN emission in the infrared is often exclusively associated with emission from the dust torus. However, it is possible for the AGN to heat dust at larger (kpc) scales. Assuming a torus opening angle of 20-70 degrees (e.g. Zhuang et al. 2018\nocite{Zhuang18}), the AGN light will heat dust in the AGN narrow line region if the black hole accretion disk is aligned with the galaxy plane (e.g. Baron et al. 2016\nocite{Baron16}), or dust in the host galaxy if the two are misaligned (e.g. Viaene et al. 2020\nocite{Viaene20}). The subsequent emission will be an additional component to the torus far-IR Rayleigh-Jeans tail.

It is currently impossible to directly measure the intrinsic or pure\footnote{Note that the definition of `intrinsic'/`pure' I adopt here, is emission exclusively linked to the AGN. This includes both direct emission from the black hole accretion disk and hot corona, but also reprocessed emission from dust heated by the AGN radiation field, i.e. dust in the vicinity of the AGN (torus) and further afield at kpc scales (narrow line region). } AGN emission over the entire electromagnetic spectrum. Nevertheless, there currently exists a variety of intrinsic AGN SEDs in the literature covering a large part of the electromagnetic spectrum, derived using a combination of data and modelling  --- e.g. Netzer et al. (2007\nocite{Netzer07}; hereafter N07); Mullaney et al. (2011\nocite{Mullaney11}, hereafter M11); Mor $\&$ Netzer (2012\nocite{MN12}, hereafter MN12); Xu et al. (2015\nocite{Xu15}; hereafter X15); Symeonidis et al. (2016, hereafter S16\nocite{Symeonidis16}); Netzer et al. (2016); Lyu $\&$ Rieke (2017, hereafter LR17\nocite{LR17}); Lani, Netzer $\&$ Lutz (2017, hereafter LNL17\nocite{LNL17}); Xu et al. (2020\nocite{Xu20}, hereafter X20). All AGN SEDs in the aforementioned works start from a similar standpoint, namely a sample of galaxies hosting AGN and the methods used to extract them can be broadly grouped into three categories: (A) SED decomposition into a star-forming and AGN component (B) computing the star-forming component first and then subtracting it from the total SED to retrieve the AGN component and (C) a combination of both. 
The AGN SEDs produced by these methods are similar in shape, with the exception of the S16 SED, which has higher far-IR/submm luminosity for a given optical/near-IR luminosity. This has provoked scrutiny by the community, resulting in concerns relating to the sample of PG QSOs used by S16, the library of star-forming galaxy SEDs employed to assign host components to the QSOs and also the method of combining these to retrieve the average AGN SED (e.g. LR17; LNL17; Stanley et al. 2018; Schulze et al. 2019; X20). 

In this paper, I aim to explain the origin of the differences between the S16 and other empirically-derived AGN SEDs; I start by summarising the derivation of the S16 SED in section \ref{sec:S16SED} and proceed to compare the S16 AGN SED to those derived by other works in sections \ref{sec:comp1}, \ref{sec:comp2} and \ref{sec:comp3}, while also addressing their criticisms of the S16 AGN SED. I present my summary and conclusions in section \ref{sec:summary}. Throughout, I adopt a concordance cosmology of H$_0$=70\,km\,s$^{-1}$Mpc$^{-1}$, $\Omega_{\rm M}$=1-$\Omega_{\rm \Lambda}$=0.3.

\section{The S16 SED}
\label{sec:S16SED}

The S16 sample consists of Palomar-Green (PG) QSOs with mid-IR spectroscopy, drawn from Shi et al. (2007; hereafter Shi07) who examined the mid-IR Spitzer/IRS spectra of PG QSOs up to z=0.5. It was restricted to $z<0.18$ in order to have a large incidence of PAH detections. This is because the 11.3$\mu$m PAH was central in determining the host contribution to the the total infrared luminosity of QSOs as it was used as a proxy for the luminosity from star-formation (8---1000$\mu$m; $L_{\rm SFIR}$). $L_{\rm SFIR}$ was computed by Shi07 as follows: for each QSO the flux of the 11.3$\mu$m PAH feature was matched with the Dale $\&$ Helou (2002\nocite{DH02}; hereafter DH02) template that gave the closest 11.3$\mu$m PAH line flux at the redshift of the object, subsequently scaling the template by $L_{\rm PAH, obj}$/$L_{\rm PAH, templ}$. As a result, the derivation of the S16 SED rests on the translation of $L_{\rm PAH}$ to a host galaxy AGN SED shape, which is in turn based on the reasonable assumption that the 11.3$\mu$m PAH is a good indicator of star-forming luminosity in AGN hosts. Indeed, PAHs have been routinely used as star formation tracers in AGN hosts (e.g. Shi et al. 2007; Lutz et al. 2008; Watabe, Kawakatu $\&$ Imanishi 2008; Rawlings et al. 2013; Esquej et al. 2014; Alonso-Herrero et al. 2014); S16 presented an extensive literature review on the topic particularly on the robustness of the 11.3$\mu$m PAH as an SFR tracer in AGN host galaxies.

S16 used the computed $L_{\rm SFIR}$ to further cull the S16 sample to the QSOs with $L_{\rm BBB} > 10 \times L_{\rm SFIR}$, where $L_{\rm BBB}$ is big blue bump luminosity (2keV-1$\mu$m; Grupe et al. 2010), in order to have a sample where the AGN is more powerful than the galaxies' stellar output. Maximizing the ratio of AGN to stellar powered emission ensured that the AGN SED could be visible over a large part of the electromagnetic spectrum, giving a larger degree of certainty in the deconvolution of the AGN emission. The final S16 QSO sample consisted of 47 optically luminous ($\nu L_{\nu 5100} > 10^{43.5}$ erg/s) unobscured and radio-quiet QSOs at $z<0.18$. 

In order to build the QSO SEDs, S16 used optical/near-IR data from Palomar, SDSS and 2MASS, mid and far-IR data from \textit{IRAS}, \textit{AKARI}, \textit{WISE}, \textit{Spitzer} and \textit{Herschel}. Subsequently, S16 assigned host SEDs to each QSO choosing the DH02 template whose $L_{\rm IR}$ was closest to the QSO's $L_{\rm SFIR}$ computed by Shi07, scaling the chosen DH02 template by $L_{\rm SFIR}$/$L_{\rm IR, templ}$. Once all QSOs had assigned host components, S16 separately averaged the host components and the QSO SEDs, subsequently subtracting the former from the latter in order to retrieve the average intrinsic AGN SED (hereafter referred to as the S16 AGN SED).

As discussed in S16 and Symeonidis (2017\nocite{Symeonidis17}; hereafter S17), the strong far-IR/sub-mm continuum characterising the S16 AGN SED suggests that powerful AGN could heat dust at kpc scales, drowning the IR emission of their host galaxy all the way up to the submm. S17 found that the average SED shape of the most luminous unobscured QSOs at $2 < z < 3.5$ was the same as the intrinsic S16 AGN SED over the entire UV-submm wavelength range. Since the S16 AGN SED was derived from a set of low redshift and intermediate luminosity QSOs, this finding plausibly indicated that the intrinsic mid/far-IR emission of AGN as a function of wavelength, is broadly independent of intrinsic AGN power [for $\nu L_{\nu, 5100}$ or $L_{\rm X (2-10 keV)}$ $>$ $10^{43.5}$ erg/s AGN] and redshift (at least up to $z \sim 3.5$), consistent with the observation that the UV-to-mid-IR SEDs of QSOs also do not evolve as a function of redshift or AGN luminosity (e.g. Hao et al. 2014). 

Fig. \ref{fig:AGNseds} shows the S16 AGN SED compared with other SEDs from the literature. These are the X15 SED (taken from LR17), the MN12 SED extended into the far-IR as described in Netzer et al. (2016; hereafter referred to as the extended MN12 SED), the M11 SEDs, the N07 SED, the X20 SED, the LNL17 SED and the LR17 SED. The common feature in all the SEDs is that they have lower far-IR/sub-mm emission than the S16 AGN SED. Part of the reason is that most were derived using methods A and C --- apart from the LR17 SED which like the S16 SED, was built using method B (see section \ref{sec:introduction} for the description of the different methods). In the sections that follow, I discuss and compare the different methods used and their impact on the characteristics of the resulting AGN SEDs.

\begin{figure}
\begin{tabular}{c}
\epsfig{file=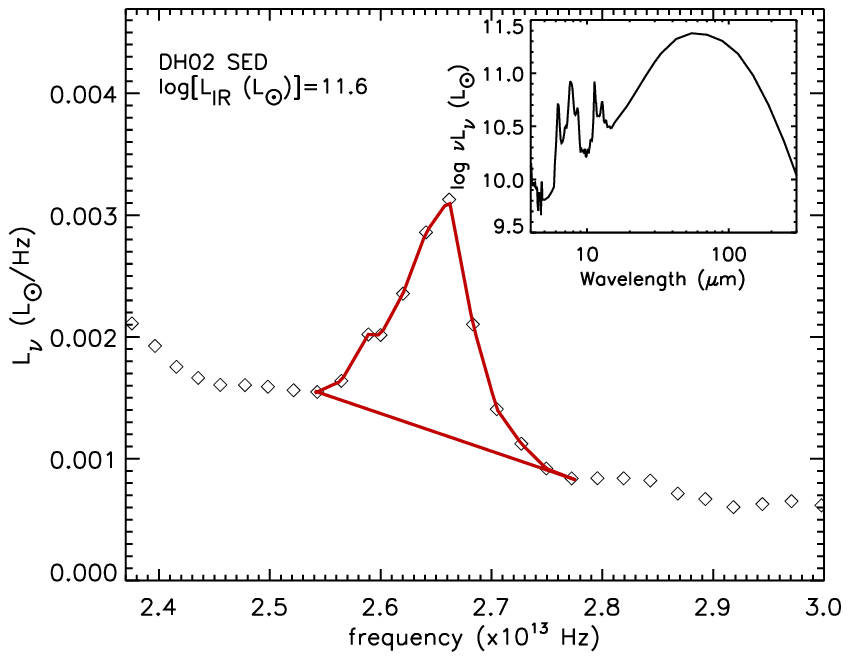,width=0.99\linewidth} \\
\epsfig{file=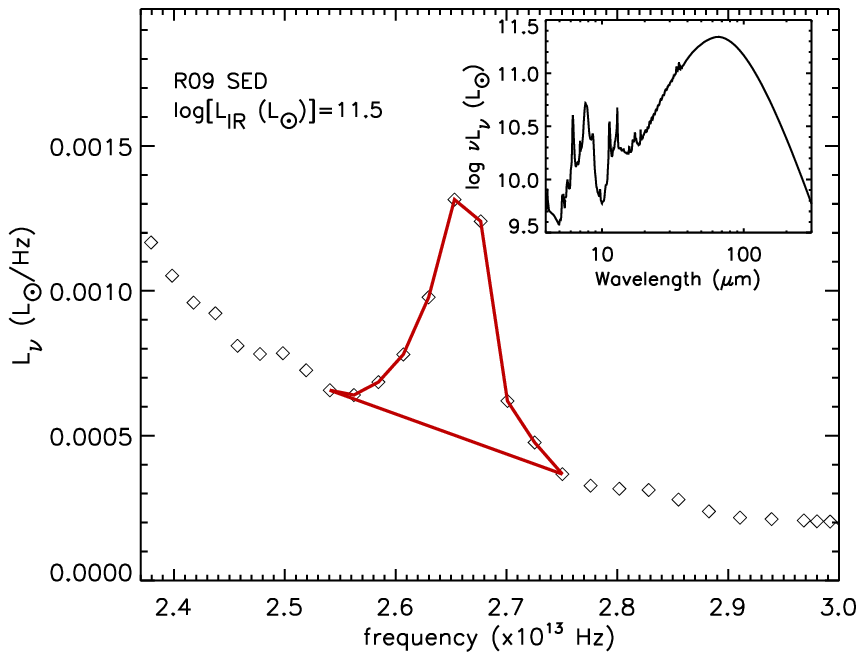,width=0.99\linewidth} \\
\end{tabular}
\caption{Example of how the 11.3$\mu$m PAH luminosity was extracted from the DH02 (top panel) and R09 (bottom panel) templates. The red outline traces the PAH and the continuum. The inset figures show the corresponding SEDs and their $L_{\rm IR}$ are quoted in the main plot. }
\label{fig:pah}
\end{figure}

\begin{figure}
\epsfig{file=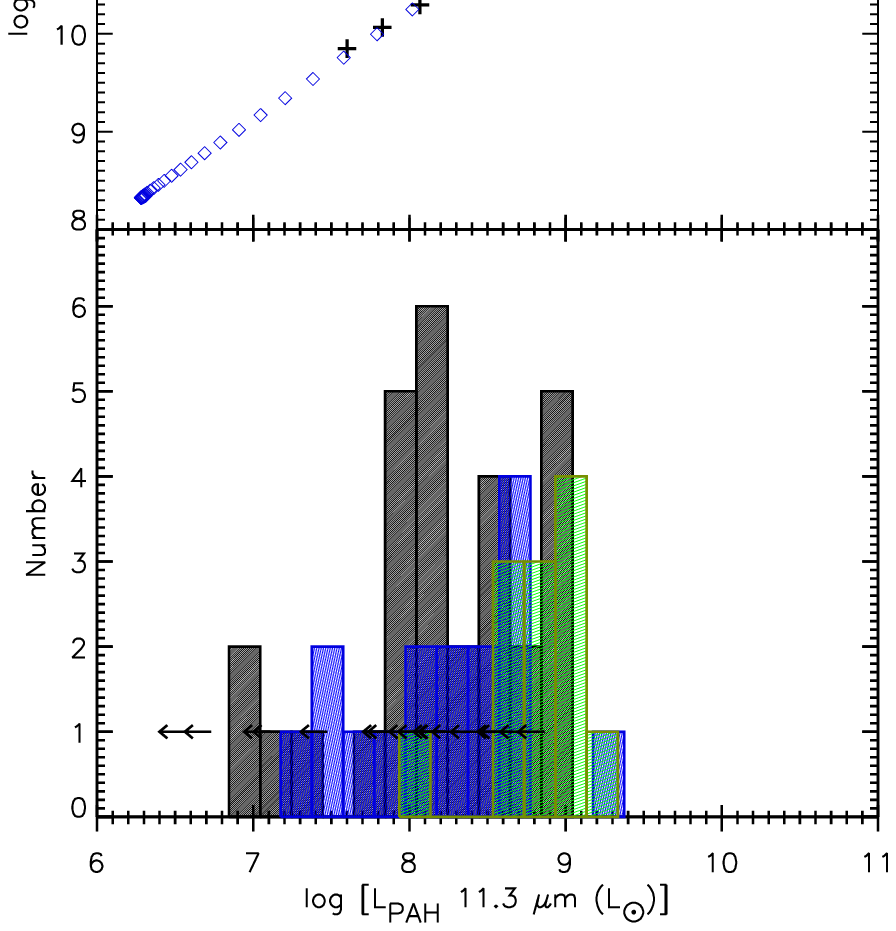,width=0.99\linewidth} 
\caption{\textit{Upper panel}: $L_{\rm IR}$ versus $L_{\rm PAH}$ (11.3$\mu$m) for the DH02 (blue diamonds) and R09 (black crosses) templates. \textit{Lower panel}: The distribution of $L_{\rm PAH}$ in the S16 sample of PG QSOs (grey histogram for PAH-detected sources; limits for PAH-undetected sources), the star-forming sample of Brandl et al. (2006; blue histogram) and the star-forming sample of Pereira-Santaella et al. (2010; green histogram)}
\label{fig:comptempl}
\end{figure}

\section{The differences between the S16 and LR17 AGN SED}
\label{sec:comp1}

As mentioned above, just like S16, LR17 use method (B) for computing the intrinsic AGN SED, where the star-forming component is computed first and then subtracted from the total SED to retrieve the AGN component. LR17 argue that the cooler nature of the S16 SED is a result of the adoption of the DH02 models to represent the host galaxy component (see section \ref{sec:S16SED}). They, on the other hand, use the Rieke et al. (2009\nocite{Rieke09}; hereafter R09) models and proceed to show that this causes the resultant AGN SED to be less luminous in the far-IR/submm than the one derived by S16 (see Fig. \ref{fig:AGNseds}). Below, I investigate the characteristics of the DH02 and R09 models in relation to the derivation of the PG QSO host galaxy components.

\subsection{The $L_{\rm IR} - L_{\rm PAH}$ parameter space covered by the DH02 and R09 SED libraries}
\label{sec:ratio}

\begin{figure}
\epsfig{file=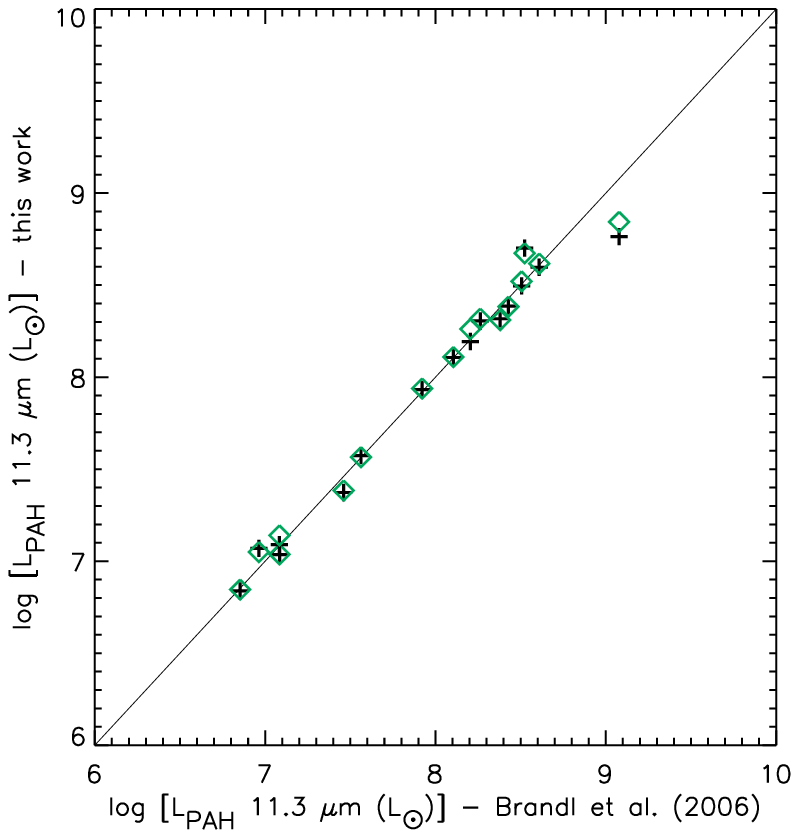,width=0.99\linewidth} 
\caption{My estimates of the 11.3$\mu$m PAH luminosity for the Br06 galaxies (linear interpolation-black crosses; spline interpolation - green diamonds) compared to the measurements of Br06. }
\label{fig:Brcomp}
\end{figure}

I examine the $L_{\rm IR} - L_{\rm PAH}$ parameter space covered by the DH02 and R09 SED libraries, using their quoted values of $L_{\rm IR}$ and computing the 11.3$\mu$m PAH luminosity for each model by integrating under the PAH in the 10.8-11.8$\mu$m range and then subtracting the linearly interpolated continuum in the same region --- see Fig \ref{fig:pah} for an example. In the top panel of Fig. \ref{fig:comptempl}, I plot $L_{\rm IR}$ versus $L_{\rm PAH}$ for both sets of SED templates. The DH02 templates were based on 69 IRAS-selected galaxies with $L_{\rm IR}$ from $10^{8}$ to $10^{12}$\,L$_{\odot}$, but excluded ULIRGs and AGN (Dale et al. 2001), whereas the R09 models were created with 9 galaxies (4 ULIRGs and 5 LIRGs). Fig. \ref{fig:comptempl} shows that the DH02 models cover a larger part of the $L_{\rm IR}$--$L_{\rm PAH}$ parameter space and are characterised by a consistent $L_{\rm IR}$/$L_{\rm PAH}$ ratio. On the other hand, the lower luminosity R09 templates do not lie on the same $L_{\rm IR}$-$L_{\rm PAH}$ relation as their higher luminosity counterparts. This seems to be in disagreement with evidence in the literature regarding the correlation between SFR and PAH luminosity --- luminous sources are seen to lie on the same relation as their lower luminosity counterparts at any given redshift (e.g. Brandl et al. 2006\nocite{Brandl06}; hereafter Br06; Pope et al. 2008\nocite{Pope08}; 2013\nocite{Pope13}), whether $L_{\rm PAH}$ is directly compared with other SFR indicators (e.g. Shipley et al. 2016; Xie $\&$ Ho 2019) or bolometric mid-IR luminosity (e.g. IRAC 8$\mu$m; a proxy for $L_{\rm PAH}$) is compared with broadband data (Wu et al. 2005; Treyer et al. 2010; Magdis et al. 2013). 

The lower panel of Fig. \ref{fig:comptempl} features the $L_{\rm PAH}$ distribution of the S16 sample of PG QSOs and the star-forming samples of Br06 and Pereira-Santaella et al. (2010\nocite{PereiraSantaella10}; hereafter PS10). One can see that the R09 library does not encompass the entire range of PAHs seen in star-forming galaxies or the S16 PG QSOs --- about 19 per cent of the S16 QSO sample is in the log\,[$L_{\rm PAH}/L_{\odot}$]\,$<7.5$ region which the R09 templates do not cover. 
As LR17 point out, applying the R09 models assigns more of the average IR emission of the QSOs to star-formation than the DH02 models. In order to examine which part of the $L_{\rm PAH}$-$L_{\rm IR}$ parameter space this difference originates from, I split the S16 PG QSO sample into three groups --- log\,[$L_{\rm PAH}/L_{\odot}$]\,$<7.5$ ($\sim$19 per cent of sources), $7.5<$log\,[$L_{\rm PAH}/L_{\odot}$]\,$<8.1$ ($\sim$28 per cent of sources) and log\,[$L_{\rm PAH}/L_{\odot}$]\,$>8.1$ ($\sim$53 per cent of sources). The groups are chosen so that the three different parts of the $L_{\rm PAH}$-$L_{\rm IR}$ parameter space are investigated separately: no overlap between the two SED libraries (group 1), overlap and good agreement (group 2) and overlap and disagreement (group 3). For each QSO, the template from each library with the closest 11.3$\mu$m $L_{\rm PAH}$ to that of the QSO is picked and subsequently scaled by $L_{\rm PAH, QSO}/L_{\rm PAH, templ}$. For all QSOs with log\,[$L_{\rm PAH}/L_{\odot}$]\,$<7.5$, the log\,[$L_{\rm PAH}/L_{\odot}$]\,$=7.5$ template is the closest one from the R09 library. The average star-forming luminosity is then computed for each QSO sub-group, separately with each SED library. I find that when using the R09 models, the star-forming luminosity ($L_{\rm SFIR}$, 8--1000$\mu$m) of the log\,[$L_{\rm PAH}/L_{\odot}$]\,$<7.5$ group, not covered by the R09 templates, is about 30 per cent higher, for the $7.5<$log\,[$L_{\rm PAH}/L_{\odot}$]\,$<8.1$ group, where the two libraries are in agreement, $L_{\rm SFIR}$ is about 6 per cent higher and for the log\,[$L_{\rm PAH}/L_{\odot}$]\,$>8.1$ group, where the two libraries diverge, it is nearly a factor of 3 higher. It is clear therefore that the main difference in $L_{\rm SFIR}$ comes from the region where the SED libraries diverge, although there is also a non-negligible contribution from the parameter space that the R09 templates do not cover.

\begin{figure}
\begin{tabular}{c}
\epsfig{file=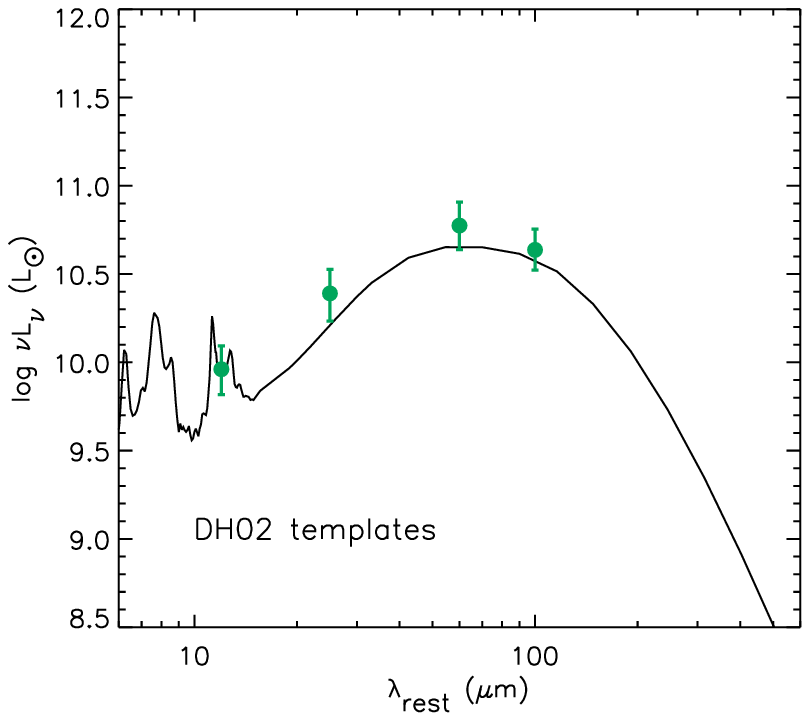,width=0.99\linewidth} \\
\epsfig{file=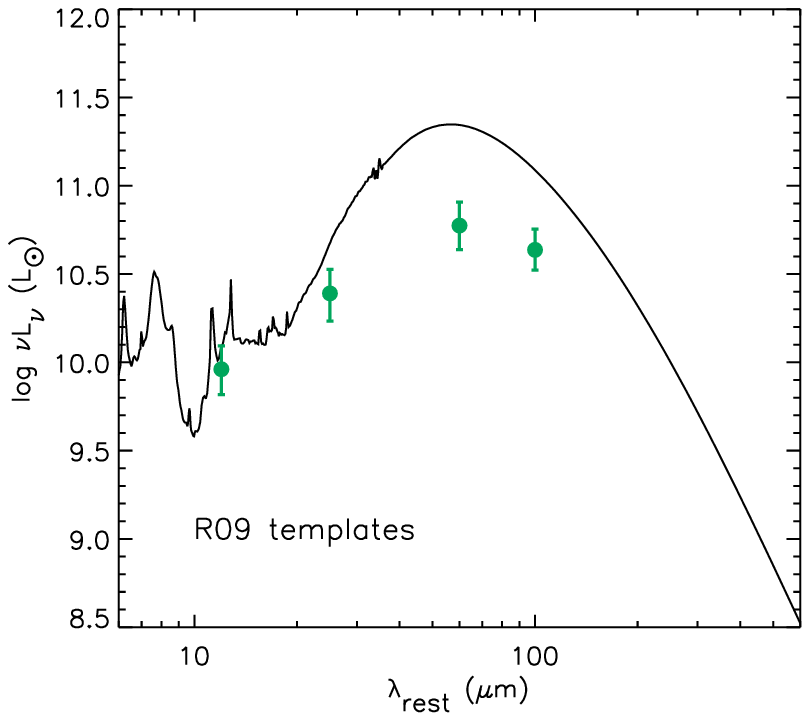,width=0.99\linewidth} \\
\end{tabular}
\caption{The average SED of star-forming galaxies (16 sources from Brandl et al. 2006): the green crosses are the monochromatic luminosities in the \textit{IRAS} 12, 25, 60 and 100$\mu$m bands. The solid curve is the average SED for this sample derived by matching the 11.3$\mu$m PAH feature of each galaxy to the DH02 models (\textit{top panel}) and the R09 models (\textit{lower panel}).}
\label{fig:seds1}
\end{figure}

\begin{figure}
\begin{tabular}{c}
\epsfig{file=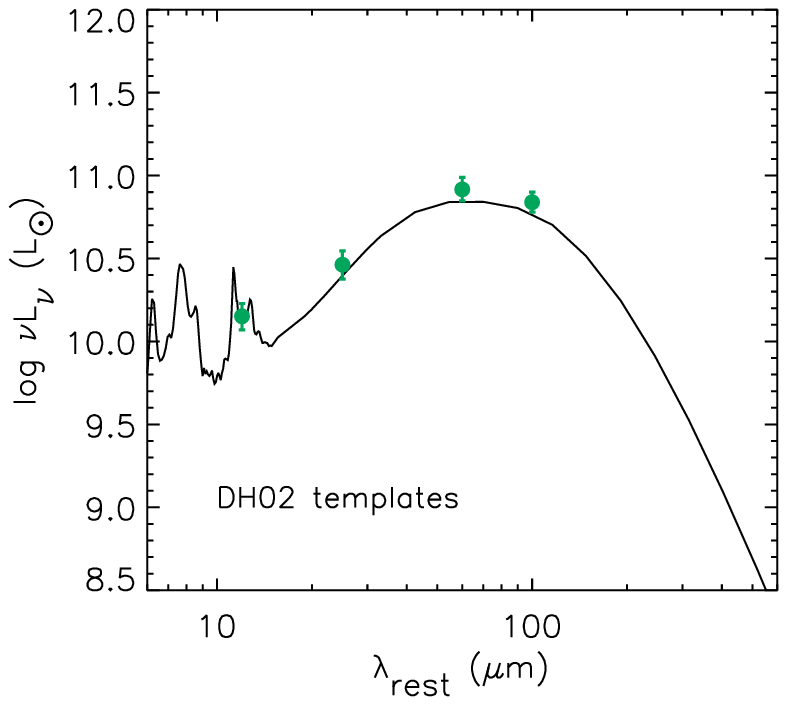,width=0.99\linewidth} \\
\epsfig{file=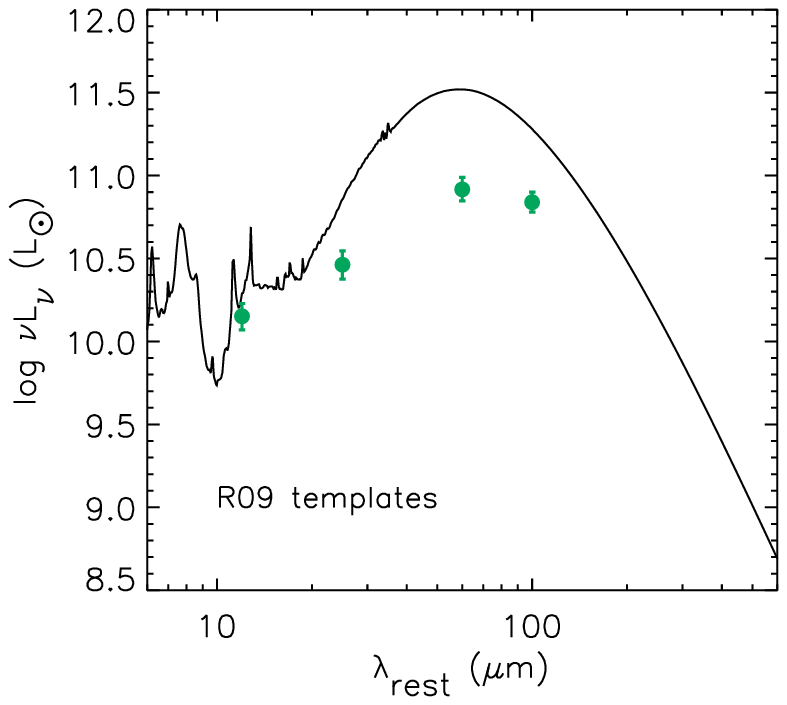,width=0.99\linewidth} \\
\end{tabular}
\caption{The average SED of star-forming galaxies (28 sources from Brandl et al. 2006 and Pereira-Santaella et al. 2010): the green crosses are the monochromatic luminosities in the \textit{IRAS} 12, 25, 60 and 100$\mu$m bands. The solid curve is the average SED for this sample derived by matching the 11.3$\mu$m PAH feature of each galaxy to the DH02 models (\textit{top panel}) and the R09 models (\textit{lower panel}).}
\label{fig:seds2}
\end{figure}

\subsection{Testing the DH02 and R09 SED libraries on star-forming galaxies}
\label{sec:test}

As described in section \ref{sec:S16SED}, the host galaxy SED for each S16 PG QSO was determined by matching the luminosity of the 11.3$\mu$m PAH to the closest $L_{\rm PAH}$ from the DH02 SED library. The assigned DH02 templates were then averaged in order to obtain the average host emission in the S16 sample of PG QSOs. S16 examined the accuracy of this method on a group of star-forming galaxies, by matching their measured 11.3$\mu$m PAHs to the DH02 models and then subsequently comparing the average matched model SED with the true average SED of the galaxies. It was shown that this method of matching the DH02 templates via the 11.3$\mu$m luminosity accurately reproduces the true average SED of star-forming galaxies. Since the main difference between the S16 and LR17 results stems from the parameter space where the DH02 and R09 templates disagree (see section \ref{sec:ratio}), here I repeat the aforementioned test in order to determine which SED library can more accurately reproduce the shape and normalisation of star-forming galaxy IR SEDs. For this test, I first use the Br06 sample of 16 star-forming galaxies and subsequently combine it with the 12 star-forming galaxies from PS10 --- see Fig. \ref{fig:comptempl} for the Br06 and PS10 11.3$\mu$m $L_{\rm PAH}$ distribution. The Br06 and PS10 samples were chosen because they consist of typical star-forming galaxies and AGN contamination in these samples is unlikely. On the other hand, the Great Observatories All-sky LIRG Survey (GOALS; Armus et al. 2009), employed by LR17 to evaluate their models, is made up of (U)LIRGs, a large fraction of which lie above the star-forming main sequence with respect to their $L_{\rm IR}/L_{\rm 8 \mu m}$ ratio (Stierwalt et al. 2014). It is not implausible that part of the reason for the boosted $L_{\rm IR}$ over 8$\mu$m emission (and by extrapolation $L_{\rm IR}/L_{\rm PAH}$ ratio) could be heavily obscured (unidentified) AGN which would contribute to the total IR luminosity, even in the subset of GOALS galaxies chosen by LR17 to be pure starbursts. Indeed, Petric et al. (2011) found a substantial AGN incidence amongst the GOALS sample based on fine structure mid-IR lines. Moreover, Koss et al. (2013), who use Swift's Burst Alert Telescope (BAT) to study the ultra-hard X-ray properties of GOALS, report that LIRGs have a higher ultra-hard X-ray detection rate than a control sample matched in redshift and stellar mass, as well as higher nuclear gas column densities than standard BAT-detected AGN. The likelihood of AGN contamination makes the GOALS sample less ideal for the subsequent test than the Br06 and PS10 samples. 

PS10 use linear interpolation for measuring PAH luminosities, as I do here for the R09 and DH02 templates (section \ref{sec:ratio}), whereas Br06 use spline interpolation. These methods are similar hence are expected to give consistent results, nevertheless I check that this is the case by measuring the PAH luminosities of the Br06 galaxies both by linear interpolation and spline interpolation. Figure \ref{fig:Brcomp} shows that my measurements are in good agreement with the Br06 published values. I subsequently match each galaxy in the Br06 sample with the DH02 template that has the closest $L_{\rm PAH}$, scaling the chosen template by $L_{\rm PAH, gal}$/$L_{\rm PAH, templ}$. I repeat the same with the R09 template library. In both cases I average the templates matched to the Br06 sample and compare with the averaged luminosities in the four \textit{IRAS} bands (12, 25, 60, 100 $\mu$m). The 1$\sigma$ uncertainties on the averaged luminosities are computed by bootstrapping with 10,000 iterations. Fig. \ref{fig:seds1} shows that the average DH02 model is close to the true average SED of the Br06 sample, whereas the mean R09 SED overpredicts the galaxies' average luminosity particularly at 60 and 100$\mu$m. I now increase the sample of star-forming galaxies available for this test, by including the sources of PS10. The results are shown in Fig. \ref{fig:seds2}. The true SED of star-forming galaxies is significantly offset from the average R09 SED, but consistent with the mean DH02 model. 

Given these results, it is reasonable to conclude that the DH02 library is the better choice for assigning the host components on the PG QSOs by matching the 11.3$\mu$m luminosity. The test specifically indicates that the $L_{\rm PAH}$/$L_{\rm IR}$ ratio characterising the R09 models is not consistent with typical star-forming galaxies, a likely consequence of the more extreme IR-luminous galaxies used to derive these models. Nevertheless, even disregarding the above test, the DH02 SEDs are arguably the better choice for deriving the host components of the PG QSOs because (i) the R09 library is more heavily weighted towards the most IR-luminous galaxies and includes ULIRGs, inconsistent with the $L_{\rm IR}$ range of the S16 QSO sample which extends down to $10^{10}\,L_{\odot}$ and does not include ULIRGs, and (ii) the $L_{\rm PAH}$ range of the S16 QSO sample extends to lower luminosities than what is covered by the R09 templates (Fig. \ref{fig:comptempl}).

\section{The differences between the S16 and LNL17 AGN SED}
\label{sec:comp2}

In section \ref{sec:comp1}, I examined the choice of star-forming SED library and how it impacts the derivation of the AGN SED, prompted by the work of LR17. Here, I investigate the argument of LNL17 that the steps that follow host component assignment to each QSO make a difference in the amount of far-IR emission characterising the AGN SED and proceed to discuss the LNL17 method of obtaining an intrinsic AGN SED. LNL17 use method (C) in their derivation of the intrinsic AGN SED (see section \ref{sec:introduction}), a hybrid of simultaneous SED decomposition and subtracting the host component in order to retrieve the AGN component. The LNL17 AGN SED is shown in Fig. \ref{fig:AGNseds}.

\subsection{The method of obtaining the AGN SED: averaging or taking the median? Normalising or not?}
\label{sec:comp2_1}

As mentioned in section \ref{sec:S16SED}, S16 separately average the host components and the QSO SEDs, subsequently subtracting the former from the latter in order to retrieve the average intrinsic AGN SED. On the other hand LNL17 subtract the corresponding host component from each QSO SED and then normalise the host-subtracted SEDs before taking the median. They propose that the difference in the two methods is the main reason behind the different intrinsic AGN SEDs produced by the S16 and LNL17 work; their argument is based on the idea that the most optically luminous QSOs have different far-IR properties compared to their lower luminosity counterparts.

I first check that individually subtracting the corresponding host component from each QSO SED and then averaging the host-subtracted SEDs (the method used by LNL17) yields the same result as averaging the QSO and host SEDs independently and subsequently subtracting the the latter from the former (the method used by S16). I find that it indeed does, since the entire process is arithmetic and hence 
\begin{equation}
\frac{1}{n}\sum\limits_{i=1}^n Q_{i} - \frac{1}{n}\sum\limits_{i=1}^n H_{i} = \frac{1}{n}\sum\limits_{i=1}^n Q_{i}- H_{i}
\end{equation}
where Q and H represent the QSO and host components respectively and i refers to each QSO.

\begin{figure}
\epsfig{file=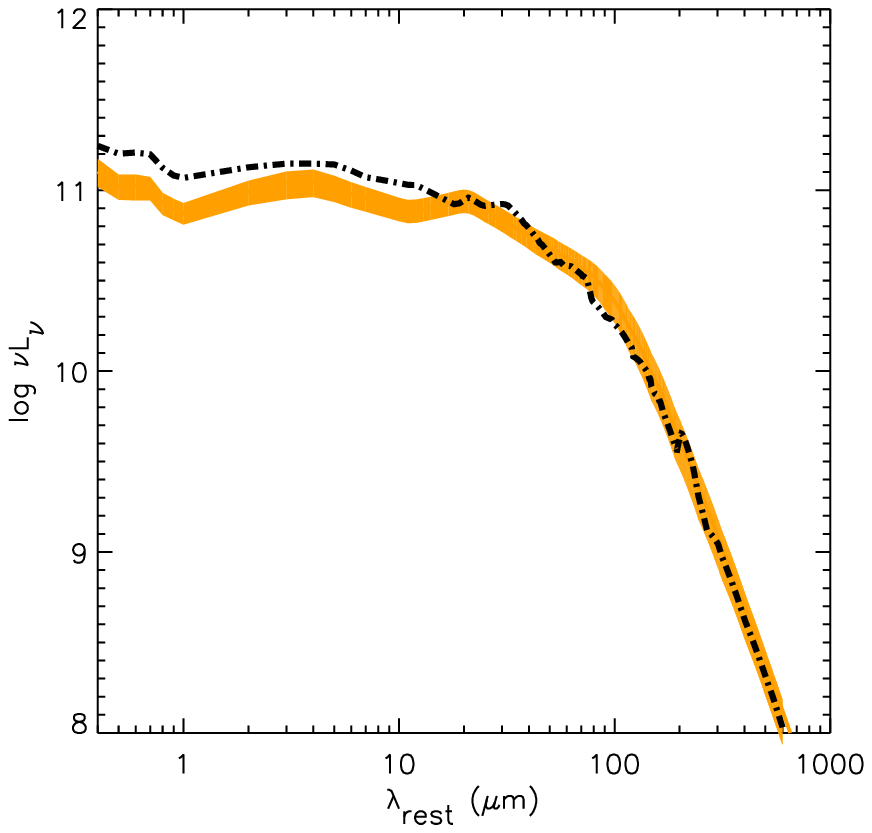,width=0.99\linewidth} 
\caption{The S16 AGN SED in yellow (the thickness indicating the 1$\sigma$ uncertainty boundaries) compared with the median host-subtracted SED from the S16 PG QSO sample. The former is computed by averaging the QSO and host SEDs separately and then subtracting to get the AGN SED. The latter is computed by subtracting the corresponding host component from each QSO SED and then taking the median of the host-subtracted SEDs. In both cases the S16 sample of PG QSOs is used. For comparing them on this plot the SEDs are shown normalised at 20$\mu$m following LNL17. }
\label{fig:residualSEDs_nofixlikeLani17_compaver}
\end{figure}

\begin{figure}
\epsfig{file=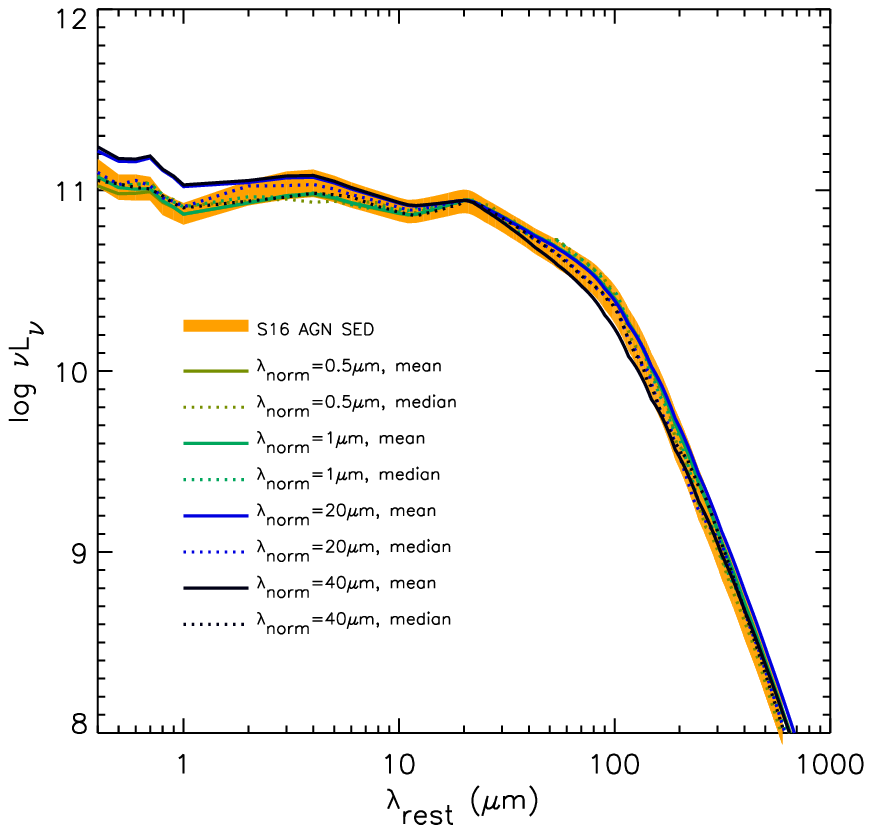,width=0.99\linewidth} 
\caption{The S16 AGN SED in yellow (the thickness indicating the 1$\sigma$ uncertainty boundaries) compared with various mean/median host-subtracted SEDs. In all cases the S16 PG QSO sample is used and the host-subtracted SEDs are normalised at various wavelengths (0.5, 1, 20, 40 $\mu$m) before taking the mean/median. For comparing them on this plot they are shown normalised at 20$\mu$m following LNL17. }
\label{fig:residualSEDs_comp}
\end{figure}

Secondly, I examine how different the median is from the mean. I re-compute the AGN SED by individually subtracting the corresponding host component from each QSO SED before taking the \textit{median} of the host-subtracted SEDs. Figure \ref{fig:residualSEDs_nofixlikeLani17_compaver} shows there is little difference between the mean and the median SEDs in the far-IR/sub-mm. 

Finally, I examine the effect of normalisation before averaging or taking the median, as follows: I subtract the host component from each QSO SED and then normalise the individual intrinsic AGN SEDs at  0.5, 1, 20 and 40 $\mu$m before averaging/taking the median. Fig. \ref{fig:residualSEDs_comp} shows that the all AGN SEDs have similar levels of far-IR/sub-mm power. Like LNL17, I also compare the 70$\mu$m/12$\mu$m colour of the average host-subtracted SEDs shown in Fig. \ref{fig:residualSEDs_comp} with that of the S16 AGN SED. I find the latter to be only about 2 per cent higher. This similarity in mid-to-far-IR colour, as well as the fact that all aforementioned ways of obtaining the AGN SED produce similar results indicate that, contrary to the LNL17 hypothesis, the most optically luminous QSOs do not have substantially different far-IR properties to their lower luminosity counterparts --- there is no apparent correlation between the AGN power in the optical and the shape of the AGN SED in the far-IR/sub-mm. This was also shown in Figure 6 of S16 where the sample was divided into two luminosity bins and the intrinsic AGN SEDs computed in each bin were compared.

\subsection{Assigning host galaxy components}
\label{sec:comp2_2}

As described in section \ref{sec:comp2_1}, the difference between the S16 and LNL17 AGN SED is unlikely to result from the method of combining the SEDs, once the host galaxy SED has been matched to each QSO. It must therefore originate at the point where each QSO is assigned a host galaxy component, or even earlier with the sample selection. 

Regarding the sample selection, the low-redshift cut-off in S16 ($z>0.18$) secured a QSO sample the majority of which had PAH 11.3$\mu$m detections. Indeed, as shown in S16, the small fraction of QSOs with no 11.3$\mu$m PAH detections meant that taking any value of 11.3$\mu$m PAH for those sources below the upper limit, changed the final AGN SED within the originally computed confidence intervals. On the other hand, the LNL17 sample is the Petric et al. (2015) sample of 85 PG QSOs targeted by \textit{Herschel}, up to $z=0.5$, and, in contrast to the S16 work, a large fraction of those (most at $z>0.18$) do not have a 11.3$\mu$m PAH detection. This likely introduces a large source of uncertainty in the LNL17 host galaxy assignment and plausibly contributes to the discrepancy between the LNL17 and S16 SEDs. The data used to build the LNL17 QSO SEDs, included \textit{Herschel}/SPIRE (250, 350, 500$\mu$m) photometry and \textit{Herschel}/PACS (160$\mu$m) photometry, which S16 also included. LNL17 additionally took advantage of 70 and 100$\mu$m data from \textit{Herschel}/PACS whereas S16 uses 70$\mu$m data from \textit{Spitzer}/MIPS and 100$\mu$m data from \textit{Herschel}/PACS. Petric et al. (2015) showed that the 70$\mu$m PACS and MIPS data are in good agreement for PG QSOs, and as a result the data used are unlikely to be a factor in the discrepancy between the S16 and LNL17 SEDs. 

I next examine the LNL17 method of assigning host galaxy components to each QSO. LNL17 employ a combination of two methods, depending on whether there is a PAH detection. For the sources with PAH detections, LNL17 use the Shi07 method of matching a DH02 model to each QSO via the 11.3$\mu$m PAH (like S16). However, for the sources without PAH detections, they employ an SED fitting method: using the extended MN12 AGN SED (shown in Fig. \ref{fig:AGNseds}) they subtract the AGN component and retrieve the SF component. This process, whereby an input AGN model is used in SED decomposition in order to retrieve the QSO host components, which are then subtracted from the QSO SEDs in order to retrieve the intrinsic AGN SEDs, suggests that the output AGN SED will inherit the shape of the input AGN model(s) used in its derivation. In other words, because the gradient of the input models is preserved in SED fitting, the output AGN SED cannot deviate outside the parameter space covered by the input AGN model(s). Consequently, for about half of the LNL17 QSO sample (the sources with no PAH detections) the final AGN SED would be similar to the input AGN SED, in this case the extended MN12 AGN SED.

In light of the above, the most plausible reason for the difference between the LNL17 and S16 AGN SEDs is the method of assigning host galaxy components to the PG QSOs. The use of the extended MN12 SED, which is itself created with the assumption of no cool dust heating by AGN (see section \ref{sec:comp3}), suggests that the LNL17 SED will have lower far-IR/sub-mm emission than the S16 AGN SED, by construction, and rather than because the data have shown it to be the case.

\section{Comparison between the S16 and other AGN SEDs}
\label{sec:comp3}

The AGN SEDs explored in this paper are shown in Fig. \ref{fig:AGNseds}. As mentioned in section \ref{sec:introduction}, empirical AGN SEDs are derived using samples of galaxies hosting AGN and can be broadly grouped into three categories: (A) SED decomposition into star-forming and AGN components (B) computing the star-forming component first and then subtracting it from the total SED to retrieve the AGN component and (C) a combination of both. Since, as also mentioned in section \ref{sec:introduction}, it is impossible to measure the intrinsic AGN SED directly, at least in the far-IR/sub-mm, methods (A) and (C) by design rely on some best-guess model for the AGN and/or star-forming component. On the other hand, galaxy SFRs (hence the host component) can be determined from observations, and as a result, it is possible for method (B) to give an independent measure of the AGN emission.

The works using methods (B) and (C), LR17 and LNL17, were explored in sections \ref{sec:comp1} and \ref{sec:comp2} respectively, where the differences between them and the S16 AGN SED were also discussed in detail. Method (A) works under the premise of a pre-defined balance between AGN and stellar power. Although this approach has been used often --- N07, M11, MN12, X15, X20 --- its crucial weakness is that the derivation of the AGN SED is strongly coupled to the input models and assumptions. For example, N07 start from the premise that most of the 50--100$\mu$m emission in PG QSOs is due to star formation, hence their resultant AGN SED is by definition constrained in the far-IR/sub-mm. Similarly, in the M11 and MN12 works, the AGN SED has been specifically built to have little far-IR/sub-mm emission. 
M11 create an input AGN model to use for SED decomposition, by adding some arbitrary amount of far-IR/sub-mm emission to AGN SEDs which are well-defined at shorter wavelengths. MN12 do something comparable: based on the assumption of no AGN heating of kpc scale dust, they create an AGN SED by simply adding a single temperature (T=100 K) modified blackbody to an already-known AGN SED at short wavelengths. On the other hand, X20 perform SED decomposition with input AGN SEDs from the Siebenmorgen et al. (2015\nocite{Siebenmorgen15}) AGN torus model library. As also described in S16, the Siebenmorgen et al. models include emission from the AGN accretion disk and torus, but do not account for kpc scale dust emission, and hence the X20 AGN SED is missing that particular component. 

The X15 SED is a modified version of the Elvis et al. (1994) AGN SED, the modification, according to X15, being a correction for star-formation. However, most of the Elvis et al. data do not probe the far-IR/sub-mm, hence the Elvis et al. QSO SED is potentially unreliable in that part of the spectrum; most of the Elvis et al. QSOs are not significantly detected at \textit{IRAS}/100$\mu$m and a large fraction ($\sim$25 per cent) are not significantly detected at \textit{IRAS}/60$\mu$m. On this premise alone, the far-IR/submm luminosity of the X15 AGN SED is likely to be uncertain. To correct the Elvis et al. SED for star-formation, X15 first linearly fit the values of the $f_{70}/f_{24}$ slope and PAH 11.3$\mu$m equivalent width (EW) for a sample of QSOs and star-forming galaxies. Subsequently, using that relation and the average PAH EW for (some of) the Elvis et al. QSOs, they compute the expected $f_{70}/f_{24}$ assuming this is equal to the factor by which the Elvis et al. AGN template has been boosted due to star-formation relative to PAH EW=0. A star-forming galaxy template is then adjusted accordingly and subtracted from the Elvis et al. AGN template. However, the rationale behind this process is not clear and it seems to be based on the assumption that in most cases far-IR emission in QSOs has a contribution from star-formation. 

In all above cases, the balance of AGN and stellar-powered dust emission is in some way pre-determined, narrowing down the parameter space that the resultant AGN SED can cover. In essence, since in all formulations the far-IR/sub-mm emission attributed to the AGN is restricted, the resultant SEDs are weaker in the far-IR/sub-mm than the S16 AGN SED, by design, and not because the data show it to be so.

\subsection{Comparing galaxy colours to AGN SEDs}

In general, empirically derived AGN SEDs, such as the ones referenced in this work, are the mean or median AGN emission of the sample they were derived from. Hence, they are ideally used on galaxy populations, i.e. compared with the average SED of galaxies in a given redshift and luminosity bin, rather than the SEDs of individual sources --- note that for the comparisons to be meaningful, the sample in each $L-z$ bin should be complete and unbiased. Nevertheless, this does not preclude their use on individual objects as long as the caveats are made clear and taken into account in the analysis. In other words, when these AGN SEDs are used on a single object, it does not necessarily represent the AGN emission of that object, rather it represents a reasonable estimate for what the AGN emission could be. 

Note that having complete $L-z$ bins for the optimum evaluation of AGN SEDs does not contradict earlier statements that the shape of intrinsic mid/far-IR emission of AGN is potentially constant as a function of AGN power and redshift. The luminosity in a $L-z$ bin refers to the total luminosity of galaxies, AGN+host, the host component potentially evolving with redshift. As a result, limiting comparisons to individual $L-z$ bins keeps the effects of star-formation at bay in order to evaluate the AGN contribution to the energy budget.

When normalised to the optical AGN power, the S16 AGN SED matches the average bolometric broadband SEDs of the most luminous unobscured QSOs at a given redshift (see S16, S17), suggesting that star-formation, or more generally dust-reprocessed stellar emission, is not necessary to power any part of their IR continuum. This is not the case for other AGN SEDs --- a stellar component is often needed to make up the shortfall in luminosity. As a result, when comparing galaxy colours against AGN SEDs, sources will scatter on either side of the S16 AGN SED if their IR emission is AGN-dominated, whereas they will lie above it if their IR emission is SF-dominated. With regard to other AGN SEDs, most sources predominantly lie above them, because their far-IR/sub-mm continuum has more power than what is described by those SEDs. In the case of SF-dominated sources, the reason is clearly an excess far-IR/sub-mm component due to star-formation. In the case of galaxies hosting AGN, while some authors claim that the reason is again far-IR/sub-mm emission from star-formation, S16 propose that it is due to AGN heating of kpc scale dust, a component which is missing in other AGN SEDs. Essentially, the S16 SED allows for the possibility that the entire IR continuum of the most luminous QSOs is AGN-powered.

\begin{figure}
\epsfig{file=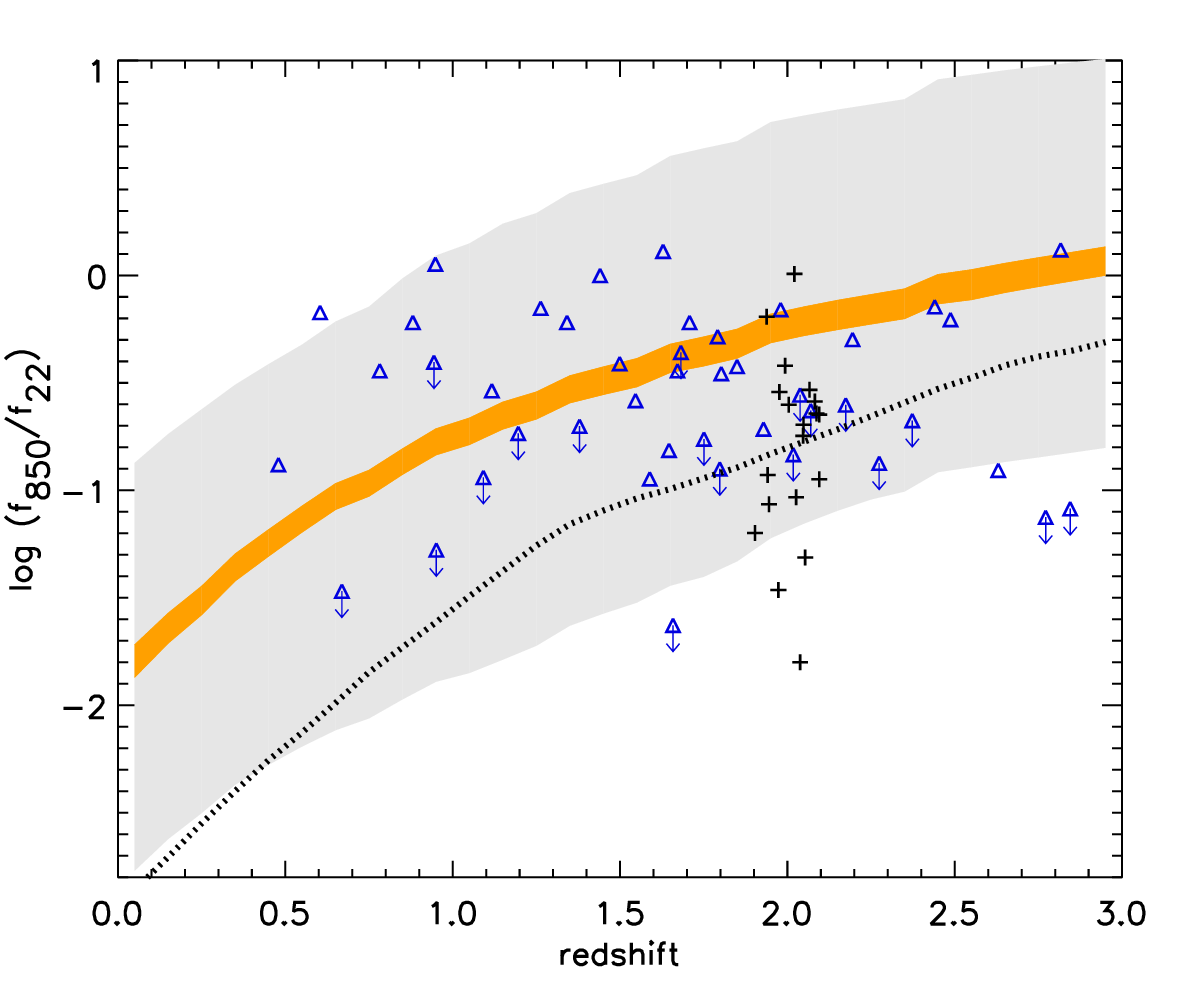,width=0.99\linewidth} 
\caption{The 850 $\mu$m to 22$\mu$m flux density ratio versus redshift for the Lonsdale et al. (2015) QSOs (blue triangles) and the Schulze et al. (2019) QSOs (black crosses). The orange curve is the parameter space covered by the S16 AGN SED with the thickness indicating the 1$\sigma$ confidence intervals. The shaded grey region represents the parameter space occupied by the individual intrinsic AGN SEDs whose average make up the S16 AGN SED (see Symeonidis 2017). The dashed black curve is the Mullaney et al. (2011) mean AGN SED. }
\label{fig:colours}
\end{figure}

With this in mind, arguments that the S16 AGN SED is inconsistent with observations because it lies above some QSOs are misconstrued. For example, LNL17 report that the median stacked luminosities of the Netzer et al. (2016) QSOs \textit{undetected} in SPIRE 250$\mu$m, lie below the S16 AGN SED. Indeed, this is where median (or mean) stacks of undetected sources should lie, since they do not represent the entire population at a given $L-z$ bin, rather only the sources with the lowest far-IR emission --- as I state above, for the comparison to be meaningful, the sample should complete and unbiased. The most luminous (log $\nu L_{\nu} (1350 \AA) > 46.7$) QSOs from Netzer et al. (2016) were in fact examined in S17. S17 showed that their average SED was consistent with the S16 AGN SED (when normalised at 1$\mu$m), with the detected and undetected sources lying above and below the S16 AGN SED respectively. 

Stanley et al. (2018) argue that the S16 AGN SED is invalid because many of the Lonsdale et al. (2015) QSOs have lower $f_{850}/f_{24}$ ratios than what is predicted by it. In Fig \ref{fig:colours}, I plot the QSO data from Lonsdale et al. (2015) against the backdrop of parameter space covered by the individual AGN SEDs which make up the average S16 AGN SED (see S17). The QSOs scatter within the AGN-dominated S16 region, suggesting that their IR emission could be AGN-dominated in the IR. The M11 mean AGN SED lies below most sources, suggesting that a star-forming component would be required in addition to an AGN component in order to reproduce their sub-mm emission. Schulze et al. (2019) also make the argument that the S16 SED is not a good representation of the AGN emission because most of the 850$\mu$m flux of their QSOs is below what is predicted by the S16 AGN SED. However, as shown in Schulze et al., their QSO sample has uncharacteristically low sub-mm emission compared to other QSO samples, which indicates that it is unlikely to be representative of typical QSOs in the particular $L-z$ range probed. Nevertheless most sources have $f_{850}/f_{22}$ colours within the parameter space covered by the S16 intrinsic AGN SEDs (Fig \ref{fig:colours}).

\section{Summary and conclusions}
\label{sec:summary}

I have investigated the difference between the S16 and other empirically-derived AGN SEDs in the literature, as well as reports of potential drawbacks in the derivation of the former. The topic of contention is the elevated far-IR/submm continuum of the S16 SED compared to other AGN SEDs. 

The derivation of the S16 AGN SED is based on the assumption that the 11.3$\mu$m PAH is a good indicator of star-forming luminosity in AGN hosts and rests on the consequent translation of $L_{\rm PAH}$ to a host galaxy SED shape. Indeed, all AGN SEDs have been derived with underlying assumptions and corresponding caveats and it is important to keep these in mind when using them. 

The LR17 SED was derived with similar assumptions to the S16 AGN SED, although LR17 used the R09 star-forming models to represent the host component, rather than the DH02 models employed by S16, leading to less pronounced far-IR/sub-mm emission in their final AGN SED. LR17 argued that the R09 models are more appropriate, however my analysis indicated that they are not the ideal choice for determining the PG QSOs' host component via PAH matching, because (i) the $L_{\rm PAH}/L_{\rm IR}$ ratio characterising the R09 models is not consistent with the far-IR SEDs of typical star-forming galaxies and (ii) they do not provide good coverage of the PG QSO $L_{\rm IR}$-$L_{\rm PAH}$ parameter space.

The assumptions and caveats characterising other AGN SEDs discussed in this work, are related to the balance of power between AGN and star-formation. In particular, a common feature amongst them is that in their formulation, it is improbable that the AGN will dominate the far-IR/sub-mm continuum of galaxies. In some cases the AGN SEDs were built by adding an arbitrary Rayleigh-Jeans drop-off to well-defined emission at shorter wavelengths. In other cases, such SEDs were subsequently used as input AGN models in SED decomposition on an object by object basis in order to retrieve a final average AGN SED, a process which results in the output AGN SED covering the same parameter space as the input AGN model(s). 

The difference in the far-IR/sub-mm continuum strength between the S16 and other AGN SEDs is also evident when comparing them against the mid-to-far-IR colours of QSOs. QSOs in a given (complete and unbiased) $L-z$ bin will scatter on either side of the S16 AGN SED if their IR emission is AGN-dominated, whereas they will lie above it if their IR emission is SF-dominated. Finding that QSO colours scatter around the S16 AGN SED does not imply that the S16 AGN SED is invalid. Instead, it indicates that the S16 SED allows for the possibility that the IR continuum of galaxies could be entirely AGN powered in the far-IR/sub-mm. On the other hand, QSOs predominantly lie above other AGN SEDs, because their far-IR/sub-mm continuum has more power than what is described by those SEDs.  Some authors claim that this is because of an additional star-forming component, whereas S16 have proposed that it is due to the component of kpc-scale dust heating by AGN missing in those SEDs.

My analysis showed that this additional far-IR/sub-mm component characterising the S16 SED is not present in other empirical AGN SEDs \textit{by construction}, and not as a consequence of erroneous steps in the derivation of the S16 SED. As a result, the proposed reasons as to why the S16 AGN SED is not a reasonable representation of AGN emission in the far-IR/sub-mm, do not hold.

\section*{Acknowledgments}
MS acknowledges support by the Science and Technology Facilities Council [ST/S000216/1].

\section*{Data Availability}
The data underlying this article are either available in the article and in its online supplementary material or will be shared on reasonable request to the corresponding author.

\bibliographystyle{mn2e}
\bibliography{references}

\end{document}